\documentclass{article}
\usepackage{spconf,amsmath,graphicx}
\usepackage{bm}
\usepackage{musicography}
\usepackage{multicol}
\usepackage{multirow}
\usepackage{booktabs}
\usepackage{xcolor}
\usepackage{url}
\usepackage{tabularx}
\usepackage{enumitem}
\newcolumntype{R}{>{\raggedleft\arraybackslash}X}
\newcolumntype{L}{>{\raggedright\arraybackslash}X}
\newcolumntype{C}{>{\centering\arraybackslash}X}

\title{Compose \& Embellish: Well-Structured Piano Performance Generation via A Two-Stage Approach}
\name{Shih-Lun Wu$^{\, \musSharp \, \musEighth}$ \qquad Yi-Hsuan Yang$^{\, \musSharp \, \musHalf}$
}
\address{
  \musSharp \;  Yating Music Team, Taiwan AI Labs, Taipei, Taiwan \\
  \musEighth \, Language Technologies Inst., School of CS, Carnegie Mellon University, Pittsburgh, PA, USA\\
  \musHalf \; Research Center for IT Innovation, Academia Sinica, Taipei, Taiwan  \\
\small{\texttt{shihlunw@andrew.cmu.edu, yhyang@ailabs.tw}}
}
\begin{document}
\ninept
\maketitle

\begin{abstract}
Even with strong sequence models like Transformers, generating expressive piano performances with long-range musical structures remains challenging.
Meanwhile, methods to compose well-structured melodies or lead sheets (melody\,$+$\,chords), i.e., simpler forms of music, gained more success.
Observing the above, we devise a two-stage Transformer-based framework that \textsc{Compose}s a lead sheet first, and then \textsc{Embellish}es it with accompaniment and expressive touches.
Such a factorization also enables pretraining on non-piano data.
Our objective and subjective experiments show that \textsc{Compose\,\&\,Embellish} shrinks the gap in structureness between a current state of the art and real performances by half, and improves other musical aspects such as richness and coherence as well.
\end{abstract}
\section{Introduction}
\label{sec:intro}
Recent years have witnessed a multitude of research works on leveraging Transformers \cite{vaswani2017attention} for symbolic music generation.
Generating piano performances emerged as a quintessential arena for such studies, for the rich musical content and texture piano playing can entail without having to deal with the complicated orchestration of instruments.
Thanks to Transformers' outstanding capability of modeling long sequences containing complex inter-token relations, generating several minutes-long expressive piano music end-to-end has been made possible \cite{huang19music, huang2020pop, hsiao21aaai}.
Though these works all claimed to have improved upon their predecessors in creating repetitive structures, a central element of music, it has been repeatedly shown that they fail to come up with overarching repetitions and musical development that hold a piece together \cite{wu2020jazz, zhang2021structure, dai2022missing}.
On the other hand, a line of research that tackles simpler forms of music, e.g., melodies or lead sheets (melody\,$+$\,chords) has seen promising results in composing well-structured pieces \cite{medeot2018structurenet, dai2021controllable, zou2022melons}.
A reasonable conjecture then follows: Could it be too demanding for a monolithic model to generate virtuosic performances end-to-end, as it has to process local nuances in texture or emotions, and  the high-level musical flow, all at once?

\begin{figure}
    \centering
    \includegraphics[width=0.8\linewidth]{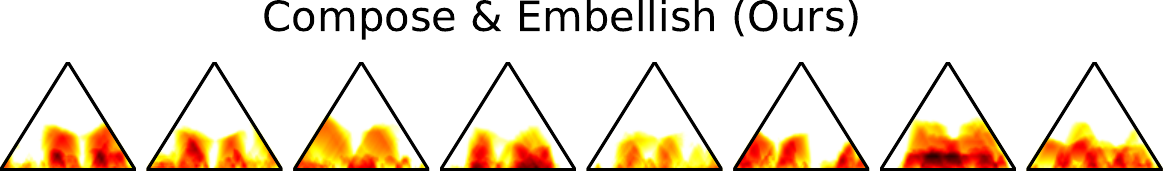} \\
    \vspace{2.5mm}
    \includegraphics[width=0.8\linewidth]{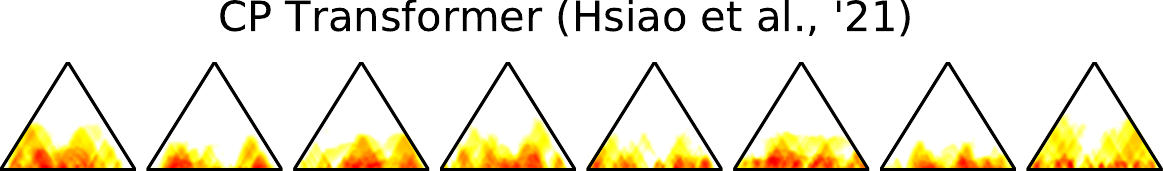} \\
    \vspace{2.5mm}
    \includegraphics[width=0.8\linewidth]{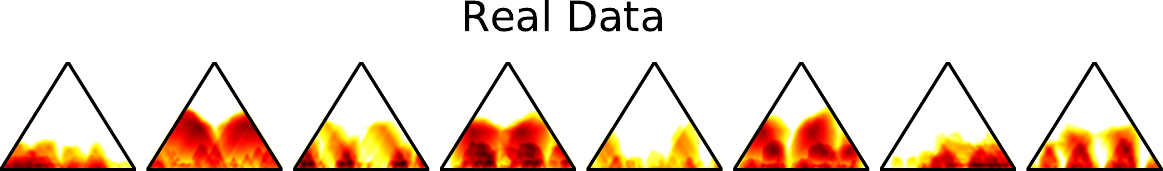}
    \vspace{-3mm}
    \caption{\textit{Fitness scape plots} \cite{muller2012scape} of pieces randomly drawn from generations by \textsc{Compose\,\&\,Embellish}, by CP Transformer \cite{hsiao21aaai}, and from real data. Darker colors towards top of the triangle indicate more significant long-range repetitive structures.}
    \label{fig:scplots}
    \vspace{-4mm}
\end{figure}

Therefore, in this paper, we split piano performance generation into two stages, and propose the \textsc{Compose\,\&\,Embellish} framework that rests on performant prior works \cite{hsiao21aaai, wu2020jazz}.
The \textsc{Compose} step writes the lead sheet that sets the overall structure of a song, while the \textsc{Embellish} step conditions on the lead sheet, and adds expressivity to it through accompaniment, dynamics, and timing.
Through experiments, we strive to answer the following research questions:
\begin{itemize}[noitemsep, topsep=0pt, leftmargin=*]
    \item \textbf{RQ \#1:} Can the two-stage framework compose better-structured piano performances than an end-to-end model, without adversely impacting diversity of musical content?
    \item \textbf{RQ \#2:} Does being able to pretrain the \textsc{Compose} step with larger amounts of non-piano data bring performance gains?
    \item \textbf{RQ \#3:} How well does the \textsc{Embellish} step follow the structure of music generated by the \textsc{Compose} step?
\end{itemize}
Fig.~\ref{fig:scplots} demonstrates our improvement over a state of the art \cite{hsiao21aaai}.
We open-source our implementation\footnote{Code: \url{github.com/slSeanWU/Compose_and_Embellish}} and trained model weights.\footnote{\url{huggingface.co/slseanwu/compose-and-embellish-pop1k7}}
Readers are encouraged to listen to samples generated by our framework.\footnote{Generated samples: \url{bit.ly/comp_embel}}

\begin{figure*}
\centering
\includegraphics[trim={5mm 50mm 2mm 42mm},clip,width=0.928\textwidth]{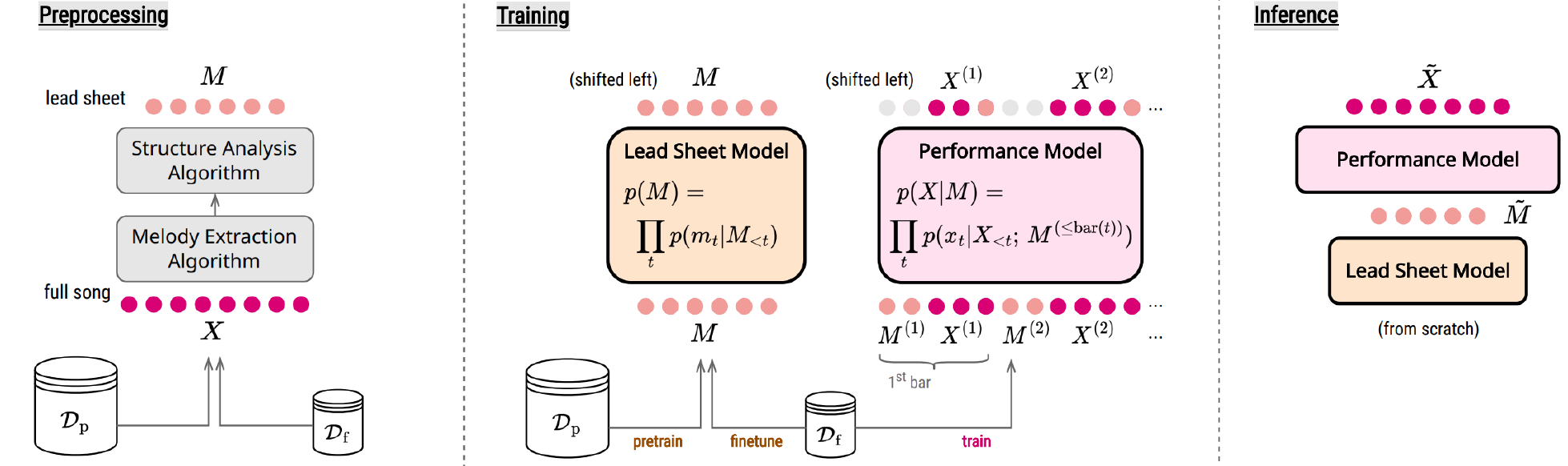}
\caption{System overview of \textsc{Compose \& Embellish}.}
\label{fig:overview}
\vspace{-2mm}
\end{figure*}

\section{Related Work}
\label{sec:rel-work}
For expressive piano performances, \cite{huang19music} and \cite{huang2020pop} showed respectively that relative positional encoding and beat-based music representation enhance generation quality.
\cite{hsiao21aaai} designed a more compact representation and utilized memory-efficient attention to fit entire performances into a Transformer.
\cite{zhang2021structure} directly addressed musical structure with a multi-granular Transformer, but weakened expressivity by abstracting timing and dynamics away.
Bar-level blueprints \cite{musemorphose21arxiv} and musical themes \cite{shih2022theme} may help to maintain long-range structure, but neither of these systems are capable of unconditioned generation.

On composing melodies or lead sheets, researchers used note-level repeat detection and modeling \cite{medeot2018structurenet}, phrase tokens \cite{wu2020jazz}, hierarchical generative pipeline \cite{dai2021controllable}, and bar-level similarity relations \cite{zou2022melons} to induce repetitive structures.
Out of these, \cite{wu2020jazz} is the most straightforward one that avoids potential error propagation between multiple components, and is hence adopted by our framework.

\section{Method}
\label{sec:method}
\subsection{Input Sequences}
\label{subsec:input}
We represent a polyphonic musical piece or performance with a sequence of tokens $X$.
To encode the expressiveness of a performance, besides chord progression and onset time/pitch/duration of notes, $X$ also contains the velocity (i.e., loudness) of each note, as well as beat-level tempo changes.
Any existing method that extracts the monophonic melody line (e.g., the \textit{skyline} algorithm \cite{uitdenbogerd1998manipulation}) may then be applied to $X$.
Using the chord progression in $X$ and the extracted melody, we additionally leverage a structure analysis  algorithm \cite{dai2020automatic} based on edit similarity and A$^{*}$ search to capture repetitive phrases (in a form like A$_1$B$_1$A$_2$...) of the piece.
The melody, chords, and structure information would constitute a \textit{lead sheet} of the piece, denoted by $M$.
Typically, $M$ is much shorter for having less notes, and the expressive aspects being discarded.
For both $X$ and $M$, there exists a mapping $\mathrm{bar}(t)$ that gives the index of the bar the $t^{\text{th}}$ token belongs to.
With the mappings, we may segment $X$ and $M$ into $\{X^{(1)},\dots,X^{(B)}\}$ and $\{M^{(1)},\dots,M^{(B)}\}$, where $B$ is the piece's number of bars.
The segmented sequences will be used in our model.

\subsection{Data Representation}
\label{subsec:data-rep}
Our token vocabulary is designed based on Revamped MIDI-derived Events (REMI) \cite{huang2020pop}.
In a full performance $X$, a \textsc{[Bar]} token appears whenever a new musical bar begins.
\textsc{[SubBeat{\textunderscore}*]} token indicate timepoints within a bar, in 16th note (\musSixteenth) resolution.
Each note is represented by three tokens: \textsc{[Pitch{\textunderscore}*]} (\textsc{A0} to \textsc{C8}), \textsc{[Duration{\textunderscore}*]} (\musSixteenth \, to \musWhole), and \textsc{[Velocity{\textunderscore}*]} (32 levels).
Moreover, \textsc{[Tempo{\textunderscore}*]} (32$\sim$224 bpm) tokens  set the pace, and  \textsc{[Chord{\textunderscore}*]} tokens (12 roots $\times$ 11 qualities) provide harmonic context.
The two above may appear as frequently as every beat.
For lead sheets $M$, we take the mean tempo, and place only one \textsc{[Tempo{\textunderscore}*]} token at the very beginning.
We also omit \textsc{[Velocity{\textunderscore}*]} of each note.
 To add structure information to $M$, we refer to \cite{wu2020jazz}:
At a phrase's starting bar, we put \textsc{[Phrase{\textunderscore}*]} (8 possible letters) and \textsc{[RepStart{\textunderscore}*]} (1$^{\text{st}}$ to 16$^{\text{th}}$ repetition) right after \textsc{[Bar]}. At the phrase's ending, \textsc{[Phrase{\textunderscore}*]} and 
\textsc{[RepEnd{\textunderscore}*]} close that bar.
An \textsc{[EOS]} token ends the entire lead sheet.
Other tokens types are the same as in $X$.
By our construction, the vocabulary size for both $X$ and $M$ is about 370.

\subsection{Models and Objectives}
\label{subsec:model}
Fig.~\ref{fig:overview} is a birds-eye view of our \textsc{Compose\,\&\,Embellish} framework.
It is made up of two generative models: the \textbf{lead sheet model} (\textsc{Compose}) $p(M)$, and the \textbf{performance model} (\textsc{Embellish}) $p(X | M)$.
While being trained independently, the two models work in tandem during inference.
For the lead sheet model, we simply factorize $p(M)$ into $\prod_t p(m_t \, | \, M_{<t})$.
It can complete a lead sheet autoregressively given a start token, i.e., \textsc{[Tempo{\textunderscore}*]}.

For $p(X | M)$, we follow the conditioned generation case in CP Transformer \cite{hsiao21aaai}, and interleave one-bar segments from $M$ and $X$ as $\{M^{(1)}, X^{(1)}, M^{(2)}, X^{(2)},\dots\}$.
This way, when generating the performance for a bar, the completed lead sheet of that bar is always the closest piece of context the model may refer to, thereby encouraging it to stay faithful to $M$.
Mathematically, $p(X | M)$ can be factorized as $\prod_t p(x_t \, | \, X_{<t}; \, M^{(\leq \mathrm{bar}(t))})$.
For the model to distinguish interleaved segments, we place \textsc{[Track{\textunderscore}$M$]} and \textsc{[Track{\textunderscore}$X$]} in front of each $M^{(\cdot)}$ and $X^{(\cdot)}$ respectively.
At inference time, we would move on to the next bar whenever \textsc{[Track{\textunderscore}$M$]} is generated.
Both models minimize the negative log-likelihood ($- \log p(\cdot)$) of the sequences.
One can use any type of sequence decoder for both models.
Due to the long sequence length (mostly $>$1k) of our data, our choice is Transformers with a causal attention mask.

Since the models are trained separately,
we may pretrain the lead sheet model on a larger amount of data ($\mathcal{D}_\mathrm{p}$ in Fig.~\ref{fig:overview}) extracted from, e.g., various multitrack pieces.
These pieces, though not played only by the piano, or at all, still likely feature a well-structured melody that is either sung or played by another instrument.
Then, we just need to finetune it on the piano performance dataset that $p(X | M)$ is trained on ($\mathcal{D}_\mathrm{f}$) to align their domain.

\section{Experiments}
\label{sec:expr}

\subsection{Datasets and Preprocessing}
\label{subsec:dset-prepro}

We adopt the full \textit{Lakh MIDI Dataset (LMD-full)} \cite{raffel2016learning} as the pretraining dataset ($\mathcal{D}_\mathrm{p}$) for our lead sheet model.
The LMD-full contains over 100k multitrack MIDIs with various instrument combinations in each of them.
The dataset for finetuning our leadsheet model and training our performance model ($\mathcal{D}_\mathrm{f}$) is \textit{Pop1K7} compiled in \cite{hsiao21aaai}.
It features about 1,700 transcribed piano performances of Western, Japanese, and Korean pop songs.

We use different algorithms to extract melodies from $\mathcal{D}_\mathrm{p}$ and $\mathcal{D}_\mathrm{f}$.
For $\mathcal{D}_\mathrm{p}$, we leverage the open-source code\footnote{\url{github.com/gulnazaki/lyrics-melody}} from \cite{melistas2021lyrics},
which searches for the instrument track whose note onset times align best with those of the song's lyrics, and regards that track as the melody.\footnote{Songs without lyrics annotations are not considered.}
For $\mathcal{D}_\mathrm{f}$, we employ the skyline algorithm \cite{uitdenbogerd1998manipulation}, which keeps only the highest-pitched note from each set of simultaneous onsets.
This simple heuristic has been shown to have a nearly 80\% accuracy on identifying pop songs' melodies \cite{chou2021midibert}.
We then perform structure analysis \cite{dai2020automatic} on the extracted melodies and discard songs with phrases spanning $>$32 bars.
The statistics of the processed datasets are displayed in Table \ref{tab:dset-stats}.
CP representation \cite{hsiao21aaai} for $\mathcal{D}_\mathrm{f}$ is made for baselining purpose.
10\% of each dataset is reserved for validation.

\begin{table}
    \footnotesize
    \centering
    \caption{Summary of datasets used in our experiments. The numbers in the last three columns are averages across a dataset.}
    \label{tab:dset-stats}
    \begin{tabularx}{\linewidth}{lc|RRRR}
    \toprule
    & \multirow{2}{*}{Repr.} & \multirow{2}{*}{\# songs} & \multirow{2}{*}{\shortstack{\# bars / \\ song}} & \multirow{2}{*}{\shortstack{\# bars / \\ phrase}} & \multirow{2}{*}{\shortstack{\# tokens \\ / song}} \\
    & & & & & \\
    \midrule
    \textit{LMD-full} ($\mathcal{D}_\mathrm{p}$) & $M$ & 14,934 & 89.6 & 8.64 & 1,356 \\
    \hline
    \multirow{3}{*}{\textit{Pop1K7} ($\mathcal{D}_\mathrm{f}$)} & $M$ & \multirow{3}{*}{1,591} & \multirow{3}{*}{104.5} & \multirow{3}{*}{8.74} & 1,841 \\
    & $X$ & & & & 5,379 \\
    & CP & & & & 2,081 \\
    \bottomrule
    \end{tabularx}
\end{table}

\begin{table*}
\footnotesize
\centering
\caption{Objective evaluation results. (All metrics are the closer to real data, the better. StDevs across individual songs follow $\pm$.)}\label{tab:obj-eval}
\begin{tabularx}{\linewidth}{L  lll | lll | lll}
\toprule
\multirow{2}{*}{} & 
\multicolumn{3}{c|}{\textbf{Structureness} (in \%)} &
\multicolumn{3}{c|}{\textbf{Melody-line Diversity}} &
\multicolumn{3}{c}{\textbf{Quality}} \\
& \multicolumn{1}{c}{$\mathcal{SI}_{\text{short}}$} & \multicolumn{1}{c}{$\mathcal{SI}_{\text{mid}}$} & \multicolumn{1}{c|}{$\mathcal{SI}_{\text{long}}$} & \multicolumn{1}{c}{$\mathcal{DN}_{\text{short}}$}
& \multicolumn{1}{c}{$\mathcal{DN}_{\text{mid}}$} & \multicolumn{1}{c|}{$\mathcal{DN}_{\text{long}}$} & \multicolumn{1}{c}{$\mathcal{H}_1$} & \multicolumn{1}{c}{$\mathcal{H}_4$} & \multicolumn{1}{c}{$\mathcal{GS}$ (\%)}  \\
\midrule
\textit{Naive repeats (1-bar)} & 83.6\scriptsize{$\,\pm$8.6} & 88.3\scriptsize{$\,\pm$5.4} & 75.7\scriptsize{$\,\pm$11} & \textcolor{white}{0}1.2\scriptsize{$\,\pm$0.6} & \textcolor{white}{0}1.2\scriptsize{$\,\pm$0.6} & \textcolor{white}{0}1.3\scriptsize{$\,\pm$0.6} & 1.95\scriptsize{$\,\pm$.40} & 1.95\scriptsize{$\,\pm$.40} & 100\scriptsize{$\,\:\pm$0.0} \\
\textit{Naive repeats (1-phrase)} & 75.5\scriptsize{$\,\pm$15} & 86.2\scriptsize{$\,\pm$6.8} & 73.9\scriptsize{$\,\pm$11} & \textcolor{white}{0}8.2\scriptsize{$\,\pm$4.8} & 10.0\scriptsize{$\,\pm$5.8} & 10.8\scriptsize{$\,\pm$6.5} & \textbf{1.96}\scriptsize{$\,\pm$.27} & \textbf{2.52}\scriptsize{$\,\pm$.22} & 83.1\scriptsize{$\,\pm$8.8} \\
\hline
CP Transformer \cite{hsiao21aaai} & 32.5\scriptsize{$\,\pm$3.3} & 29.9\scriptsize{$\,\pm$4.4} & 17.9\scriptsize{$\,\pm$6.3} & 82.0\scriptsize{$\,\pm$8.9} & 99.6\scriptsize{$\,\pm$0.7} & 100\scriptsize{$\,\:\pm$0.0} & 1.73\scriptsize{$\,\pm$.20} & 2.45\scriptsize{$\,\pm$.11} & 82.6\scriptsize{$\,\pm$9.3} \\
\hline
\textsc{Compose\,\&\,Embellish} & \textbf{36.8}\scriptsize{$\,\pm$6.7} & \textbf{35.1}\scriptsize{$\,\pm$7.7} & \textbf{25.8}\scriptsize{$\,\pm$12} & \textbf{49.7}\scriptsize{$\,\pm$19} & 69.9\scriptsize{$\,\pm$20} & 81.6\scriptsize{$\,\pm$17} & 1.99\scriptsize{$\,\pm$.17} & \textbf{2.54}\scriptsize{$\,\pm$.17} & \textbf{81.0}\scriptsize{$\,\pm$8.4} \\
\quad \textit{w/o} struct & \textbf{36.8}\scriptsize{$\,\pm$6.8} & 34.2\scriptsize{$\,\pm$9.2} & 23.8\scriptsize{$\,\pm$11} & 48.0\scriptsize{$\,\pm$17} & 68.7\scriptsize{$\,\pm$17} & 82.7\scriptsize{$\,\pm$14} & 1.99\scriptsize{$\,\pm$.19} & 2.57\scriptsize{$\,\pm$.15} & 82.1\scriptsize{$\,\pm$8.5} \\
\quad \textit{w/o} pretrain & 36.6\scriptsize{$\,\pm$7.5} & 33.1\scriptsize{$\,\pm$8.8} & 19.6\scriptsize{$\,\pm$10} & 53.2\scriptsize{$\,\pm$19} & \textbf{74.9}\scriptsize{$\,\pm$18} & \textbf{87.9}\scriptsize{$\,\pm$14} & 1.97\scriptsize{$\,\pm$.22} & 2.49\scriptsize{$\,\pm$.19} & 81.7\scriptsize{$\,\pm$9.6} \\
\quad \textit{w/o} struct\,\&\,pretrain & 36.3\scriptsize{$\,\pm$6.0} & 34.1\scriptsize{$\,\pm$6.7} & 23.0\scriptsize{$\,\pm$8.4} & 52.5\scriptsize{$\,\pm$18} & 76.1\scriptsize{$\,\pm$16} & 89.0\scriptsize{$\,\pm$11} & 2.00\scriptsize{$\,\pm$.22} & 2.57\scriptsize{$\,\pm$.18} & 82.7\scriptsize{$\,\pm$8.7} \\
\hline
Real data & 43.8\scriptsize{$\,\pm$7.1} & 43.1\scriptsize{$\,\pm$8.4} & 34.8\scriptsize{$\,\pm$12} & 50.0\scriptsize{$\,\pm$14} & 74.1\scriptsize{$\,\pm$14} & 88.3\scriptsize{$\,\pm$11} & 1.96\scriptsize{$\,\pm$.22} & 2.53\scriptsize{$\,\pm$.15} & 75.5\scriptsize{$\,\pm$8.5} \\
\bottomrule
\end{tabularx}
\vspace{-4mm}
\end{table*}

\subsection{Model Implementation Details}
\label{subsec:model-impl}
Our lead sheet model $p(M)$ is parameterized by a 12-layer Transformer \cite{vaswani2017attention} (512 hidden state dim., 8 attention heads, 41 million trainable parameters) with relative positional encoding proposed in \cite{dai2019transformer}.
With a batch size of 2, we can set maximum sequence length to 2,400 (longer than 98\% \& 90\% of songs in $\mathcal{D}_\mathrm{p}$ \& $\mathcal{D}_\mathrm{f}$, respectively) on an RTX3090-Ti GPU with 24G memory.
We use Adam optimizer with 200 steps of learning rate warmup to a peak 
of 1e$-$4, followed by 500k steps of cosine decay.
We keep a checkpoint after each epoch, finding that checkpoints with around 0.4 training NLL produces outputs of the best perceived quality.
Pretraining on $\mathcal{D}_\mathrm{p}$ requires over 5 days, while finetuning on $\mathcal{D}_\mathrm{f}$ takes less than half a day.

The performance model $p(X|M)$ is a 12-layer linear Transformer with Performer \cite{choromanski2021rethinking} attention (38 mil.~trainable parameters).
In each epoch, a random 3,072 token-long crop of interleaved $M$ and $X$ segments (see Sec.~\ref{subsec:model} for explanation) of each song is fed to the with batch size$\,=\,$4.
This sequence length corresponds to roughly 40 bars of performance.
While its possible to feed full performances using batch size$\,=\,$1, we observe that the increased attention overhead and reduced batch size render training slow and unstable.
Training the performance model takes around 3 days.\footnote{Other model hyperparameters, hardware, optimizer settings, and checkpoint selection criterion are the same as those for the lead sheet model.}

Nucleus sampling \cite{holtzman2019curious} with temperatured softmax is employed during inference.
We discover that the temperature $\tau$ and probability mass truncation point $p$ greatly affect the intra-sequence repetitiveness and diversity of generated lead sheets.
Thus, we follow \cite{fu2021theoretical} and search within $\tau=\{1.2, 1.3, 1.4\}$ and $p=\{.95, .97, .98, .99\}$ to find a combination with which our lead sheet model generates outputs with the closest mean perplexity (measured by the model itself) to that of validation real data.
Finally, $\tau = 1.2$ and $p=.97$ are chosen.
The effects of these two hyperparameters on the performance model are more subtle.
We pick $\tau=1.1$ and $p=.99$ for they lead to the most pleasant sounding performances to our ears.

\subsection{Baselines and Ablations}
\label{subsec:baseline}
We select Compound Word (CP) Transformer \cite{hsiao21aaai} as our baseline, for 
it represents the state of the art end-to-end model for unconditional expressive piano performance generation that has access to the full context of previously generated tokens, due to its low memory footprint.
However, this advantage does not lead to well-structured generations, as pointed out by \cite{zou2022melons}.
For our \textsc{Compose \& Embellish} framework, to examine whether (1) pretraining on $\mathcal{D}_\mathrm{p}$, and (2) adding structure (phrase) tokens contribute to its success,
we put to test three ablated versions without either, or both, of them.\footnote{Sampling hyperparameters ($\tau$ \& $p$) for the ablated models are chosen in the same way as described in Sec.~\ref{subsec:model-impl}.}

Additionally, we sample single-bar, and single-phrase, excerpts from the real data and repeat such excerpts to the song's length (in \# bars), to see how \textit{naive repetitions} would compare to the models.

\begin{table}
\footnotesize
\centering
\caption{Difference in structureness indicator scores between lead sheets and full performances.}\label{tab:diff-si}
\begin{tabularx}{\linewidth}{l RR|RR|RR}
\toprule
& \multicolumn{2}{c|}{Lead sheet} & \multicolumn{2}{c|}{Performance} & \multicolumn{2}{c}{$\Delta$} \\
& $\mathcal{SI}_{\text{mid}}$ & $\mathcal{SI}_{\text{long}}$ & $\mathcal{SI}_{\text{mid}}$ & $\mathcal{SI}_{\text{long}}$ & $\mathcal{SI}_{\text{mid}}$ & $\mathcal{SI}_{\text{long}}$ \\
\midrule
\textsc{C\&E} & 42.9 & 36.6 & 35.1 & 25.8 & \textcolor{white}{0}$-$7.8 & $-$10.8 \\
\; \textit{w/o} struct & 44.4 & 36.5 & 34.2 & 23.8 & $-$10.2 & $-$12.7 \\
\; \textit{w/o} pretrain & 41.1 & 28.9 & 33.1 & 19.6 & \textcolor{white}{0}$-$8.0 & \textcolor{white}{0}$-$9.3 \\
\; \textit{w/o} both & 41.3 & 31.3 & 34.1 & 23.0 & \textcolor{white}{0}$-$7.2 & \textcolor{white}{0}$-$8.3 \\
\hline
Real data & 43.6 & 36.6 & 43.1 & 34.8 & \textcolor{white}{0}$-$0.5 & \textcolor{white}{0}$-$1.8 \\
\bottomrule
\end{tabularx}
\vspace{-3mm}
\end{table}

\begin{table}
\centering
\footnotesize
\caption{User study MOS results. (\textbf{C}{\scriptsize o}\textbf{h}{\scriptsize erence}, \textbf{C}{\scriptsize o}\textbf{r}{\scriptsize rectness}, \textbf{S}{\scriptsize tructureness}, \textbf{R}{\scriptsize ichness}, \textbf{O}{\scriptsize verall}. SDs across individual data points follow $\pm$.)}\label{tab:user-study}
\begin{tabularx}{\linewidth}{l CCCCC}
\toprule
& \textbf{Ch} & \textbf{Cr} & \textbf{S} & \textbf{R} & \textbf{O} \\
\midrule
CPT \cite{hsiao21aaai} & 2.38\scriptsize{$\pm$0.9} & 2.49\scriptsize{$\pm$0.9} & 2.33\scriptsize{$\pm$0.9} & 2.64\scriptsize{$\pm$0.9} & 2.33\scriptsize{$\pm$0.9} \\
\textsc{C\&E} (ours) & 3.53\scriptsize{$\pm$0.9} & 3.11\scriptsize{$\pm$1.0} & 3.36\scriptsize{$\pm$1.2} & 3.29\scriptsize{$\pm$1.0} & 3.18\scriptsize{$\pm$0.9} \\
Real data & 4.42\scriptsize{$\pm$0.7} & 4.13\scriptsize{$\pm$0.8} & 4.44\scriptsize{$\pm$0.8} & 4.24\scriptsize{$\pm$0.8} & 4.40\scriptsize{$\pm$0.7} \\
\bottomrule
\end{tabularx}
\vspace{-3mm}
\end{table}

\subsection{Objective Evaluation}
\label{subsec:obj-eval}
We utilize a set of metrics that can be computed on a song's notes, or its synthesized audio, to evaluate the intra-song structureness, diversity, and general quality of the generated music.
\begin{itemize}[itemsep=0pt, topsep=0pt, leftmargin=*]
    \item \textbf{Structureness Indicators ($\mathcal{SI}$):} proposed first by \cite{wu2020jazz}, this metric takes the maximum value from a specific timescale range (e.g., all 10$\sim$20 seconds long segments) of the fitness scape plot \cite{muller2012scape} (see Fig.~\ref{fig:scplots} for examples) computed on the audio of a song.
    This value represents the extent to which the most salient repetitive segment in that timescale range is repeated throughout the entire song.
    We set the timescale ranges to \textbf{4}$\sim$\textbf{12}, \textbf{12}$\sim$\textbf{32}, and \textbf{over 32} seconds to capture the short-, medium-, and long-term structureness (denoted as $\mathcal{SI}_{\text{short}}$, $\mathcal{SI}_{\text{mid}}$, and $\mathcal{SI}_{\text{long}}$) respectively.
    \item \textbf{Percentage of Distinct Pitch N-grams in Melody ($\mathcal{DN}$):} following the popular \textit{dist-n} \cite{li2016diversity} metric used in natural language generation to evaluate the diversity of generated content, we compute the percentage of distinct n-grams in the pitch sequence of the skyline extracted from each full performance.
    We regard \textbf{3$\sim$5}, \textbf{6$\sim$10}, and \textbf{11$\sim$20} contiguous notes as short, medium, and long excerpts, and compute $\mathcal{DN}_{\text{short}}$, $\mathcal{DN}_{\text{mid}}$, and $\mathcal{DN}_{\text{long}}$ accordingly.
    \item \textbf{Pitch Class Histogram Entropy ($\mathcal{H}_1$, $\mathcal{H}_4$):} proposed in \cite{wu2020jazz} to see if a model uses primarily a few pitch classes (i.e., \textsc{C}, \textsc{C\#},..., \textsc{B\musFlat}, \textsc{B}, 12 in total), or more, and more evenly, of them, hence leading to a higher harmonic diversity.
    The subscripts denote whether histograms are accumulated over 1-, or 4-bar segments.
    \item \textbf{Grooving Similarity ($\mathcal{GS}$)}: used also in \cite{wu2020jazz}. It calculates the pairwise similarity of each bar's groove vector $\bm{g}$ (binary, indicating which sub-beats have onsets) as $1 - \mathrm{HammingDistance}(\bm{g}_a, \bm{g}_b)$.
    All bar pairs $(a,b)$ 
    are involved, not just adjacent ones.
    Higher $\mathcal{GS}$ suggests  
    more consistent rhythm patterns across the song.
\end{itemize}

We generate 100 songs with each model to compute the metrics.
Due to high computation cost of fitness scape plots, we only sample 200 songs from real data ($\mathcal{D}_\mathrm{f}$) for comparison.

\subsection{User Study}
\label{subsec:user-study}
We recruit 15 subjects who are able to spend around half an hour to take part in our listening test.
Each test taker is given three independent sets of music.
There are three piano performances (full song, about 3$\sim$5 minutes long each) in each set, composed respectively by (1) a human composer, (2) \textsc{Compose\,\&\,Embellish}, and (3) CP Transformer.
To facilitate comparison, the three performances in a set share the same 8-bar prompt drawn from our validation split.
Test takers are asked to rate each performance on the 5-point Likert scale, on the following aspects:
\begin{itemize}[itemsep=0pt, topsep=1pt, leftmargin=*]
\item \textbf{Coherence (Ch):} Does the music follow the prompt well, and unfold smoothly throughout the piece?
\item \textbf{Correctness (Cr):} Is the music free of inharmonious notes, unnatural rhythms, and awkward phrasing?
\item \textbf{Structureness (S):} Are recurring motifs / phrases / sections, and reasonable musical development present?
\item \textbf{Richness (R):} Is the music intriguing and full of variations within?
\item \textbf{Overall (O):} Subjectively, how much do you like the music?
\end{itemize}

\section{Results and Discussion}
\label{sec:results}
We run the metrics described in Sec.~\ref{subsec:obj-eval} on all our model variants and baselines.
The results are shown in Table \ref{tab:obj-eval}.
First and foremost, we compare CP Transformer (abbr.~as CPT henceforth) to the full \textsc{Compose\,\&\,Embellish} (\textsc{C\&E}).
On $\mathcal{SI}$ metrics, \textsc{C\&E} sits right in the middle of CPT and real data, with its advantage over CPT increasing as the timescale goes up.
Although CPT scores high on $\mathcal{DN}$'s, we should keep in mind that it is not trained to take extra care of the melody, and hence likely does not know some melodic content should be repeated.
On the other hand, $\mathcal{DN}$ scores of \textsc{C\&E} are close to those of real data, and a lot better than the two excessively repetitive baselines.
An affirmative answer to our \textbf{RQ \#1} may be given: 
\textsc{C\&E} composes better-structured piano performances without sacrificing musical diversity within a piece.
Worthwhile to note is that, despite the high $\mathcal{DN}$'s, CPT gets considerably lower $\mathcal{H}_1, \mathcal{H}_4$, and slightly higher $\mathcal{GS}$ (vs.~\textsc{C\&E} and real data), suggesting that its music may actually sound bland harmonically and rhythmically.

Next, we pay attention to the full vs.~ablated versions of \textsc{Compose\,\&\,Embellish}.
From Table \ref{tab:obj-eval}, we may observe that variants not pretrained on $\mathcal{D}_\mathrm{p}$ suffer losses on $\mathcal{SI}_{\text{mid}}$ and $\mathcal{SI}_{\text{long}}$.
Somewhat to our surprise is that \textit{w/o pretrain} performs worse than \textit{w/o struct\,\&\,pretrain}.
A possible explanation is that the introduction of structure-related tokens renders a larger amount training data necessary, as the concept of long-range repetition those tokens carry is less explicit than direct interactions between notes.
Whether this reasoning holds, however, warrants further study.
Despite worse longer-range $\mathcal{SI}$ scores and higher $\mathcal{DN}$ scores, we discover that, contrarily, the w/o pretrain variants are more prone to \emph{over-repetition} (which we define to be: one bar of melody being exactly and consecutively repeated over 6 times, or two neighboring bars of melody  repeated over 4 times)---5.5\% of these two variants' generations suffer from it, while only 2.5\% of pieces by pretrained variants and 0.5\% of real data do.
We may now answer another \textit{yes} to our \textbf{RQ \#2}: pretraining the lead sheet model helps with not only the structureness, but also the quality consistency, of generated music.

Table \ref{tab:diff-si} displays how much longer-range structureness slip after we feed generated lead sheets to our performance model.
The performance model falls far short of real performances in terms of structureness, regardless of whose lead sheets it conditions on, while our best lead sheet model already performs  structure-wise similarly to real data.
(Lead sheet $\mathcal{SI}_{\text{long}}$ of the full \textsc{C\&E} model is significantly higher than that of the w/o pretrain variants with $p < .01$, reaffirming our answer to \textbf{RQ \#2}.)
To get a better sense our performance model's issues, we check the \emph{melody matchness} \cite{hsiao21aaai} it achieves (i.e., the percentage of notes in a melody that a performance model copies and pastes into the performance; ideally 100\%)---it gets $>$98\% on lead sheets by all four variants.
Hence, a reasonable response to our \textbf{RQ \#3} would be: The performance model follows the melody faithfully, but some aspects of repetitive structures come inherently with the accompaniment and expressive details, which cannot be captured even with effective melody conditioning.

The mean opinion scores (MOS) obtained from our user study is listed in Table \ref{tab:user-study}.
As expected, \textsc{C\&E} holds significant advantage over CPT on all five aspects ($p < .01$ on 45 sets of comparisons).
Coherence (\textbf{Ch}) and structureness (\textbf{S}) are what our model does particularly well, gaining $>$1 points over CPT.
This indicates that explicit modeling of lead sheets helps \textsc{C\&E} better glue generated music into one piece, as well as reuse and develop musical content.
Nonetheless, the scores also corroborate ($p<.01$) that, by every criterion, our model still has a long way to go to rival real performances.

\section{Conclusion}
\label{sec:conclusion}
In this paper, we have introduced a two-stage framework, \textsc{Compose \& Embellish}, to generate piano performances with lead sheets as the intermediate output.
Promising prior art \cite{hsiao21aaai, wu2020jazz} were chosen and integrated to form our model backbone.
We showed via objective and subjective study that our framework composes better-structured and higher-quality piano performances compared to an end-to-end model.
Furthermore, pretraining the 1$^{\text{st}}$-stage (i.e., lead sheet) model with extra data contributed to a sizable performance gain.
Future endeavors may focus on redesigning the performance model to further close the gap between generated and real performances.

\vfill\pagebreak

\bibliographystyle{IEEEbib}
\bibliography{refs}

\begin{thebibliography}{10}

\bibitem{vaswani2017attention}
Ashish Vaswani, Noam Shazeer, Niki Parmar, Jakob Uszkoreit, Llion Jones,
  Aidan~N Gomez, Lukasz Kaiser, and Illia Polosukhin,
\newblock ``Attention is all you need,''
\newblock in {\em Proc. NeurIPS}, 2017.

\bibitem{huang19music}
Cheng{-}Zhi~Anna Huang, Ashish Vaswani, Jakob Uszkoreit, Ian Simon, Curtis
  Hawthorne, Noam Shazeer, Andrew~M. Dai, Matthew~D. Hoffman, Monica
  Dinculescu, and Douglas Eck,
\newblock ``Music {T}ransformer: Generating music with long-term structure,''
\newblock in {\em Proc. {ICLR}}, 2019.

\bibitem{huang2020pop}
Yu-Siang Huang and Yi-Hsuan Yang,
\newblock ``{Pop Music Transformer}: {Generating} music with rhythm and
  harmony,''
\newblock in {\em Proc. ACM Multimedia}, 2020.

\bibitem{hsiao21aaai}
Wen-Yi Hsiao, Jen-Yu Liu, Yin-Cheng Yeh, and Yi-Hsuan Yang,
\newblock ``{Compound Word Transformer}: {Learning} to compose full-song music
  over dynamic directed hypergraphs,''
\newblock in {\em Proc. AAAI}, 2021.

\bibitem{wu2020jazz}
Shih-Lun Wu and Yi-Hsuan Yang,
\newblock ``The {J}azz {T}ransformer on the front line: Exploring the
  shortcomings of {AI}-composed music through quantitative measures,''
\newblock in {\em Proc. ISMIR}, 2020.

\bibitem{zhang2021structure}
Xueyao Zhang, Jinchao Zhang, Yao Qiu, Li~Wang, and Jie Zhou,
\newblock ``Structure-enhanced pop music generation via harmony-aware
  learning,''
\newblock in {\em Proc. ACM Multimedia}, 2021.

\bibitem{dai2022missing}
Shuqi Dai, Huiran Yu, and Roger~B Dannenberg,
\newblock ``What is missing in deep music generation? a study of repetition and
  structure in popular music,''
\newblock in {\em Proc. ISMIR}, 2022.

\bibitem{medeot2018structurenet}
Gabriele Medeot, Srikanth Cherla, Katerina Kosta, Matt McVicar, Samer Abdallah,
  Marco Selvi, Ed~Newton-Rex, and Kevin Webster,
\newblock ``{StructureNet}: Inducing structure in generated melodies,''
\newblock in {\em Proc. ISMIR}, 2018.

\bibitem{dai2021controllable}
Shuqi Dai, Zeyu Jin, Celso Gomes, and Roger~B Dannenberg,
\newblock ``Controllable deep melody generation via hierarchical music
  structure representation,''
\newblock in {\em Proc. ISMIR}, 2021.

\bibitem{zou2022melons}
Yi~Zou, Pei Zou, Yi~Zhao, Kaixiang Zhang, Ran Zhang, and Xiaorui Wang,
\newblock ``Melons: generating melody with long-term structure using
  transformers and structure graph,''
\newblock in {\em Proc. ICASSP}, 2022.

\bibitem{muller2012scape}
Meinard M{\"u}ller and Nanzhu Jiang,
\newblock ``A scape plot representation for visualizing repetitive structures
  of music recordings.,''
\newblock in {\em Proc. ISMIR}, 2012.

\bibitem{musemorphose21arxiv}
Shih-Lun Wu and Yi-Hsuan Yang,
\newblock ``{MuseMorphose}: Full-song and fine-grained music style transfer
  with one {Transformer VAE},''
\newblock {\em arXiv preprint arXiv:2105.04090}, 2021.

\bibitem{shih2022theme}
Yi-Jen Shih, Shih-Lun Wu, Frank Zalkow, Meinard Muller, and Yi-Hsuan Yang,
\newblock ``{Theme Transformer}: Symbolic music generation with
  theme-conditioned {Transformer},''
\newblock {\em IEEE Transactions on Multimedia}, 2022.

\bibitem{uitdenbogerd1998manipulation}
Alexandra~L Uitdenbogerd and Justin Zobel,
\newblock ``Manipulation of music for melody matching,''
\newblock in {\em Proc. ACM Multimedia}, 1998.

\bibitem{dai2020automatic}
Shuqi Dai, Huan Zhang, and Roger~B Dannenberg,
\newblock ``Automatic analysis and influence of hierarchical structure on
  melody, rhythm and harmony in popular music,''
\newblock in {\em Proc. Joint Conf. AI Music Creativity}, 2020.

\bibitem{raffel2016learning}
Colin Raffel,
\newblock {\em Learning-Based Methods for Comparing Sequences, with
  Applications to Audio-to-MIDI Alignment and Matching},
\newblock Ph.D. thesis, Columbia University, 2016.

\bibitem{melistas2021lyrics}
Thomas Melistas, Theodoros Giannakopoulos, and Georgios Paraskevopoulos,
\newblock ``Lyrics and vocal melody generation conditioned on accompaniment,''
\newblock in {\em Proc. Workshop on NLP for Music and Spoken Audio (NLP4MusA)},
  2021.

\bibitem{chou2021midibert}
Yi-Hui Chou, I-Chun Chen, Chin-Jui Chang, Joann Ching, and Yi-Hsuan Yang,
\newblock ``{MidiBERT-piano}: large-scale pre-training for symbolic music
  understanding,''
\newblock {\em arXiv preprint arXiv:2107.05223}, 2021.

\bibitem{dai2019transformer}
Zihang Dai, Zhilin Yang, Yiming Yang, Jaime~G Carbonell, Quoc Le, and Ruslan
  Salakhutdinov,
\newblock ``Transformer-{XL}: Attentive language models beyond a fixed-length
  context,''
\newblock in {\em Proc. ACL}, 2019.

\bibitem{choromanski2021rethinking}
Krzysztof Choromanski, Valerii Likhosherstov, David Dohan, Xingyou Song,
  Andreea Gane, Tamas Sarlos, Peter Hawkins, Jared Davis, Afroz Mohiuddin,
  Lukasz Kaiser, David Belanger, Lucy~J Colwell, and Adrian Weller,
\newblock ``Rethinking attention with {Performers},''
\newblock in {\em Proc. ICLR}, 2021.

\bibitem{holtzman2019curious}
Ari Holtzman, Jan Buys, Li~Du, Maxwell Forbes, and Yejin Choi,
\newblock ``The curious case of neural text degeneration,''
\newblock in {\em Proc. ICLR}, 2019.

\bibitem{fu2021theoretical}
Zihao Fu, Wai Lam, Anthony Man-Cho So, and Bei Shi,
\newblock ``A theoretical analysis of the repetition problem in text
  generation,''
\newblock in {\em Proc. AAAI}, 2021.

\bibitem{li2016diversity}
Jiwei Li, Michel Galley, Chris Brockett, Jianfeng Gao, and Bill Dolan,
\newblock ``A diversity-promoting objective function for neural conversation
  models,''
\newblock in {\em Proc. NAACL}, 2016.

\end{thebibliography}

\end{document}